\documentclass[10pt]{article}

\usepackage{amsmath}
\usepackage{graphicx}
\usepackage{geometry}
\usepackage{tikz}
\usepackage{amssymb}
\usepackage{caption}
\usepackage{authblk}
\usepackage[marginal]{footmisc}

\title{Traffic flow phase transition phenomena based on the kinetic approach \footnote{\noindent This work is supported by National Key R\&D Program of China (Nos. 2021YFA1000300 and 2021YFA1000302) and National Natural Science Foundation of China (No. 12331014). \newline Email address: 51215500035@stu.ecnu.edu.cn (Z. Zhang), chlu@math.ecnu.edu.cn (C. Lu)} }

\author[a]{Zhizhen Zhang}
\author[b]{Changhong Lu}
\affil[a,b]{\small School of Mathematical Sciences, Key Laboratory of MEA(Ministry of Education) \& Shanghai Key Laboratory of PMMP, East China Normal University, Shanghai 200241, China}

\begin{document}

\maketitle

\begin{abstract}
We develop a discrete Boltzmann-type model that uses dynamics in phase space to describe the behavior of traffic flows. Firstly we model the traffic flow at mesoscopic scale using dynamics in phase space, which is considered as an evolution model in a graph, and we demonstrate the existence of phase transition phenomena through theoretical analysis. Secondly we assumed the density of vehicles in geometric space to be homogeneous and single-peaked, respectively, and performed numerical simulations to obtain the results consistent with experience. According to this model, we can perform effective simulations on the effects played by multiple parameters on the traffic flow.

\end{abstract}

\section{Introduction}

The study of traffic flow dynamics has gained increasing attention in recent years due to its impact on socioeconomic development and people's daily life. Over the past decades, researchers have proposed various theoretical models at different scales to represent the complexity of traffic systems. Microscopic models simulate vehicle interactions\cite{1,2}, mesoscopic models apply kinetic theory\cite{3,4}, and macroscopic models analyze system-level behavior\cite{5,6}.

Pioneering work by Prigogine $et$ $al.$ in the 1960s modeled traffic as a dissipative particle system using a Boltzmann-type kinetic equation\cite{a}. This represented the transport of vehicles in velocity space according to collisions and acceleration/deceleration processes. This allows us to use statistical physics to study the dynamic properties of traffic flow. However, some have argued this approach lacked foundations in observed driver behavior\cite{b}. Later studies sought to enhance mesoscopic representations by incorporating factors like following dynamics, desired speeds based on conditions, and vehicle characteristics\cite{c,d,e}. Continuous efforts aim to develop multiscalar models integrating insights across scales while balancing theoretical accuracy with computational feasibility. Overall, improved theoretical traffic models remain an active area of research with practical implications.

In experience, the traffic flow on a roadway shifts between different phases of free flow and congestion depending on the traffic density. In this paper, the study focus on phase transition analysis based on Boltzmann-type kinetic approach. In the second part we will introduce the classical Prigogine model, and the evolution of our discrete Boltzmann model of traffic flow through the graph, which incorporates the microscopic interactions between vehicles and followers. Then we analyze the phase transition phenomena in spatially homogeneous and heterogeneous states. In the third part we perform numerical simulations and analysis, and in the fourth part we have the conclusion and foresights.

\section{Model Description}

The Boltzmann-type equation describes the probability $f^{t}(x,v)$ that a vehicle is at position $x$ at moment $t$ with velocity $v$. So the entire phase space is a two-dimensional space $(x,v)$, and $f^{t}(x,v)dvdx$ represents how many vehicles are in the small cell $[v,v+dv]\times [x,x+dx]$. So once we are clear about the evolution of the probability distribution function, we can get the dynamics of the whole system.

\subsection{Prigogine model}
Prigogine $et$ $al.$ proposed the kinetics models for traffic flow\cite{a}. They suggested that the traffic process can be described by a relaxation term $Q_{\mathrm{rel}}$ and an interaction term $Q_{\mathrm{int}}$, where the relaxation term captures the expectation of acceleration toward a desired speed and the interaction term captures the obstructive effect of vehicle interaction. So the model can be written as
\begin{equation}\label{1}
  \frac{\partial f}{\partial t}+\frac{\partial (fv)}{\partial x} = Q_{\mathrm{rel}}+Q_{\mathrm{int}}.
\end{equation}
Assume the desired velocity distribution is $f^{0}(x,v)$ and the relaxation time is $\tau$ which describe how long to reach the desired speed. So the relaxation term $Q_{\mathrm{rel}}$ can be written as
\[
Q_{\mathrm{rel}} = \left(\frac{\partial f}{\partial t}\right)_{\mathrm{rel}} = \frac{f^{t}(x,v)-f^{0}(x,v)}{\tau}.
\]
We assume that $v<v^{'}$. The interaction rule is that for a given position $x$, it will change from $f^{t}(x,v^{'})$ to $f^{t}(x,v)$ with probability $f^{t}(x,v)f^{t}(x,v^{'})$ at the next moment. So the interaction item can be written as
\begin{equation*}
 \begin{aligned}
      Q_{\mathrm{int}} &=\left(\frac{\partial f}{\partial t}\right)_{\mathrm{int}}\\
       & =f^{t}(x,v)\int_{v}^{\infty}dw(1-p)(w-v)f^{t}(x,w)-f^{t}(x,v)\int_{0}^{v}dw(1-p)(v-w)f^{t}(x,w)\\
       & = (1-p)f^{t}(x,v)\int_{0}^{\infty}dw(w-v)f^{t}(x,w)\\
       & = (1-p)\rho(V-v)f,
    \end{aligned}
\end{equation*}
  where $p$ denotes the probability that a slower vehicle can be overtaken, $V$ denotes the average speed, $\rho$ denotes the vehicle density, $q$ denotes the traffic flux. And the relationship holds:
  \[
  \rho(x,t)=\int dv f^{t}(x,v),\quad q(x,t)=\int dv v f^{t}(x,v).
  \]

  \subsection{Discretized graphical evolutionary models}

For the theoretical analysis of the Bolzmann-type traffic flow model, the more common analytical methods include spatial homogenization and turning it into a weak form to obtain macroscopic information\cite{f}.
Some studies also approximate the traffic process as a combination of fast and slow processes\cite{f,g}.

We find it natural to discretize such PDE models into dynamical models of graphs to simulate this kind of jumping behavior. For a graph $G=(V,E)$, $V$ denotes the vertex set and $E$ denotes the edge set. Here we can define a directed graph with $(V_{\max}+1)\times (X_{\max}+1)$ vertices to simulate the evolutionary process. Note that here $V_{\max}$ should be a function of the total road density $\rho_{\mathrm{sum}}$, i.e. $V_{\max}=V_{\max}(\rho_{\mathrm{sum}})$. Let $V_{x,v}$ represents the state as velocity $v$ with position $x$. The vertex weights $f^{t}(x,v)$ represent the probability distribution function of the corresponding state $(x,v)$ at that moment in time. And the edge weights $W(i,j)$ represent the probability that there is a jump from vertex $i$ to vertex $j$ in the next moment $(W(i,j)\neq W(j,i))$. Assume that the adjacency matrix of the graph is $A$, then We gain the process
\begin{equation}\label{2}
  \rho^{t+1}=\rho^{t}A,
\end{equation}
where $\rho^{t} = \rho_{\mathrm{sum}}\{f^{t}(0,0),\dots, f^{t}(0,V_{\max}),f^{t}(1,0),\dots, f^{t}(X_{\max},V_{\max})\}$.
It needs to note that $f$ denotes the probability of the distribution and $\rho_{\mathrm{sum}}f$ denotes the specific density distribution.

  Due to the lack of realism in the continuous model, we intend to show more features consistent with traffic flow in the discrete dynamic model. We refer to the interaction rules for metacellular automata. In a way, on the one hand we can consider this as a good approximation of the continuous PDE equations, and on the other hand as a generalized theoretical analysis on the basis of cellular automata. We assume that the following interaction rules:
  \begin{itemize}
    \item 1. Randomly diffuse with probability $p$, i.e. with probability $p$ speed minus $1$.
    \item 2. If a fast vehicle encounters a slow vehicle at the same location, the following behavior occurs, i.e., assume $v<w$, then with probability $f^{t}(x,v)f^{t}(x,w)$ decelerates from $f^{t}(x,w)$ to $f^{t}(x,v)$ at the next moment.
    \item 3. The remainder that does not diffuse and interact accelerates $1$ with probability $q$, and with probability $(1-q)$ maintains the original speed.
  \end{itemize}

We expect phase transition phenomena to arise spontaneously during this evolution, i.e., for different initial densities, the evolution of the graph should eventually converge to different stable points. But immovable points exist in $v$-space and not in $(x,v)$-space, so we need to get the dynamics of $v$-space.

\begin{equation}\label{3}
  f^{t+1}(v)=H^{t}_{v+1}+H^{t}_{\tau\ge v}+H^{t}_{v-1}+H^{t}_{v}.
\end{equation}
where

\begin{equation}\label{4}
  H^{t}_{v+1}=\left\{
  \begin{aligned}
  &pf^{t}(v+1), &\quad 0\le v\le V_{\max}-1\\
  &0,&\quad v=V_{\max}.
  \end{aligned}
  \right
.
\end{equation}
\begin{equation}\label{5}
  H^{t}_{\tau\ge v}=\left\{
  \begin{aligned}
  &\sum_{x=0}^{X_{\max}}\left[f^{t}(x,0)f^{t}(x,0)+\sum_{\tau=1}^{V_{\max}}(1-p)f^{t}(x,\tau)f^{t}(x,0)\right], &v=0\\
  &\sum_{x=0}^{X_{\max}}\left[\sum_{\tau=v}^{V_{\max}}(1-p)^2f^{t}(x,\tau)f^{t}(x,v)\right] ,&1\le v\le V_{\max}.
  \end{aligned}
  \right
.
\end{equation}
\begin{equation}\label{6}
\footnotesize
  H^{t}_{v-1}=\left\{
  \begin{aligned}
  &0,& v=0\\
  &q\left[f^{t}(0)-\sum_{x=0}^{X_{\max}}f^{t}(x,0)f^{t}(x,0)\right] ,&v=1\\
  &q\left[(1-p)f^{t}(v-1)-\sum_{x=0}^{X_{\max}}\left((1-p)f^{t}(x,v-1)f^{t}(x,0)+\sum_{\tau=1}^{v-1}(1-p)^2f^{t}(x,\tau)f^{t}(x,v-1)\right)\right], &2\le v\le V_{\max}
  \end{aligned}
  \right
.
\end{equation}
\begin{equation}\label{7}
\footnotesize
  H^{t}_{v}=\left\{
  \begin{aligned}
  &(1-q)[f^{t}(0)-\sum_{x=0}^{X_{\max}}f^{t}(x,0)^2], & v=0\\
  &(1-q)\left[(1-p)f^{t}(v)-\sum_{x=0}^{X_{\max}}\left((1-p)f^{t}(x,v)f^{t}(x,0)+\sum_{\tau=1}^{v}(1-p)^2f^{t}(x,\tau)f^{t}(x,v)\right)\right], & 1\le v\le V_{\max}-1\\
  &(1-q)\left[f^{t}(v)-\sum_{x=0}^{X_{\max}}\left((1-p)f^{t}(x,v)f^{t}(x,0)+\sum_{\tau=1}^{v}(1-p)^2f^{t}(x,\tau)f^{t}(x,v)\right)\right],  & v=V_{\max}\\
  \end{aligned}
  \right
.
\end{equation}

\subsection{Dynamics in the phase space}

We will introduce two treatments that will eventually transform the probability distribution in $(x,v)$-space into a probability distribution in $v$-space.

Assume $\rho_{\mathrm{sum}}(0\le \rho_{\mathrm{sum}}\le 1)$ denotes the total vehicle density, satisfies:
\[
\rho_{\mathrm{sum}}=\sum_{v}f(v)\rho_{\mathrm{sum}}.
\]
$D_{+}(v)$ denotes the probability that vehicle with current speed $v$ will have a speed greater than $v$ in the next moment, $D_{-}(v)$ denotes the probability that vehicle with current speed $v$ will have a speed lower than $v$ in the next moment.

Figure 1 is a schematic.
\begin{figure}
  \centering
  \includegraphics[width=8cm]{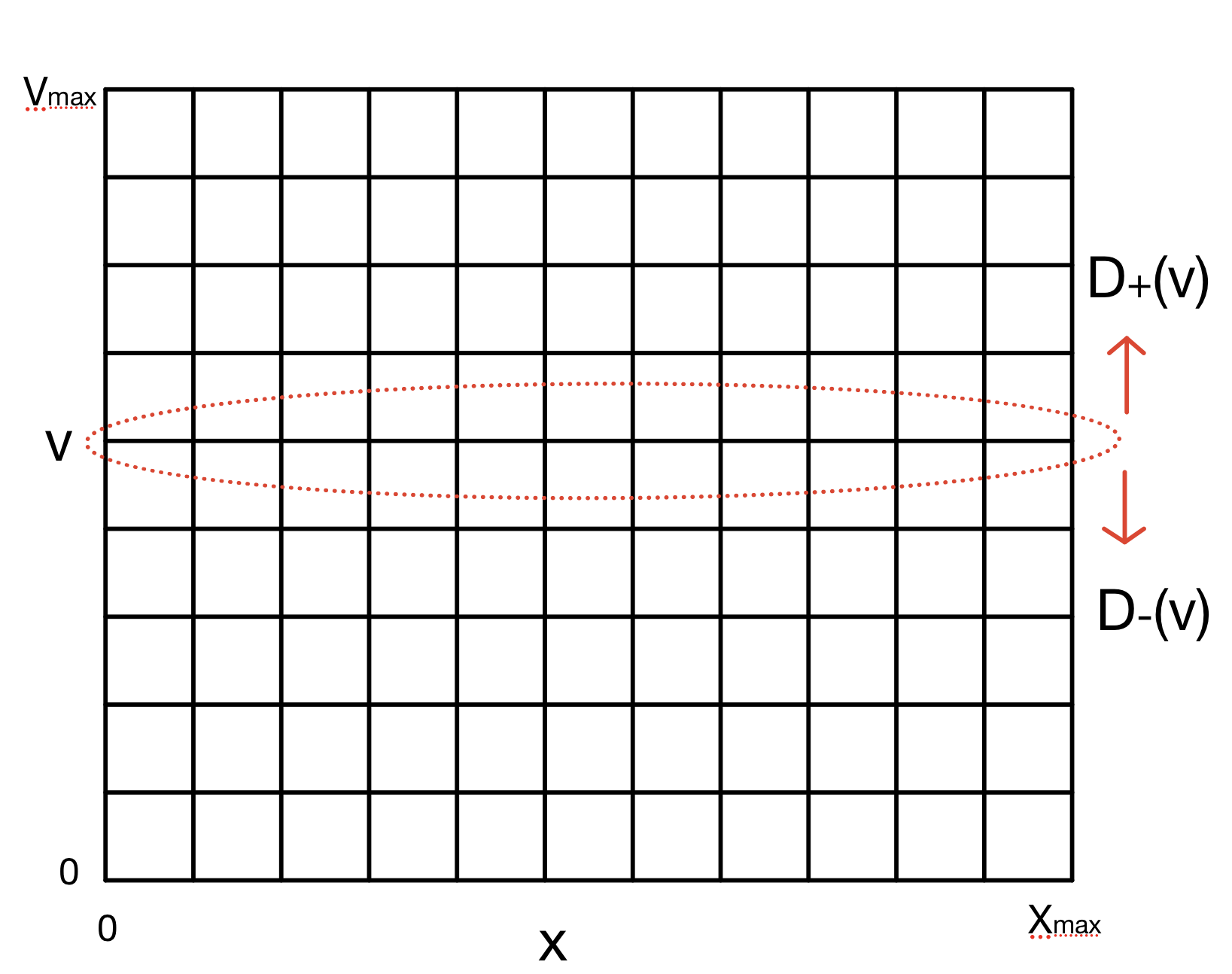}
  \caption{Schematic}\label{1}
\end{figure}

For computational simplicity, we assume the existence of a decelerating diffusion for $v=0$ and an acceleration for $V_{\max}$. We get the difference between the distribution of all accelerating and decelerating cars at the next moment as
\begin{equation}\label{8}
\begin{aligned}
A-D&=\sum_{v}\left(D_{+}(v)-D_{-}(v)\right)\rho_{\mathrm{sum}}\\
&=\sum_{v}q\left((1-p)f^{t}(v)\rho_{\mathrm{sum}}-\sum_{x=0}^{X_{\max}}\left(\sum_{\tau=0}^{v}(1-p)^2f^{t}(x,v)f^{t}(x,\tau)\rho^{2}_{\mathrm{sum}}\right)\right)\\
&-\sum_{v}\left(pf^{t}(v)\rho_{\mathrm{sum}}+\sum_{x=0}^{X_{\max}}\left(\sum_{\tau=0}^{v}(1-p)^2f^{t}(x,v)f^{t}(x,\tau)\rho^{2}_{\mathrm{sum}}\right)\right)\\
&=\left(q(1-p)-p\right)\rho_{\mathrm{sum}}-(1+q)\sum_{v}\left(\sum_{x=0}^{X_{\max}}\left(\sum_{\tau=0}^{v}(1-p)^2f^{t}(x,v)f^{t}(x,\tau)\rho^{2}_{\mathrm{sum}}\right)\right)
\end{aligned}
\end{equation}

The probability of acceleration is at most $q(1-p)$, the probability of deceleration is at least $p$. So if $q(1-p)<=p$, it is clear that the distribution of $f(v)$ will gradually converge to $V_{\max}$. In the following let's move on to the case of $q(1-p)>p$.

\subsubsection{Spatial homogenization}

Spatial homogenization is a common means of approximation. In this case:
\begin{equation}\label{9}
  f^{t}(x,v)=\frac{f^{t}(v)}{X_{\max}+1}
\end{equation}

Combining \ref{8} and \ref{9}, we obtain
\begin{equation}\label{10}
A-D=\left(q(1-p)-p\right)\rho_{\mathrm{sum}}-(1+q)\sum_{v}\left(\sum_{\tau=0}^{v}(1-p)^2f^{t}(v)f^{t}(\tau)\frac{\rho^{2}_{\mathrm{sum}}}{X_{\max}+1}\right).
\end{equation}
\begin{equation*}
  \begin{aligned}
  A-D&\le \left(q(1-p)-p\right)\rho_{\mathrm{sum}}- (1+q)(1-p)^2\sum_{v}\sum_{\tau=0}^{v}f^{t}(\tau)f^{t}(v)\frac{\rho^{2}_{\mathrm{sum}}}{X_{\max}+1},\\
  &\le \left(q(1-p)-p\right)\rho_{\mathrm{sum}}-(1+q)(1-p)^2\frac{\rho^{2}_{\mathrm{sum}}}{2(X_{\max}+1)}.
  \end{aligned}
\end{equation*}
So when
\[
\rho_{\mathrm{sum}}> \frac{q(1-p)-p}{(1+q)(1-p)^2}2(X_{\max}+1),
\]
The whole system will have more cars accelerating in all $t$ cases. We can say that in this case the traffic flow will tend to be congested.

Similarly we can have
\begin{equation*}
  \begin{aligned}
  A-D&\ge \left(q(1-p)-p\right)\rho_{\mathrm{sum}}- (1+q)(1-p)^2\sum_{v}\sum_{\tau=0}^{v}f^{t}(\tau)f^{t}(v)\frac{\rho^{2}_{\mathrm{sum}}}{X_{\max}+1},\\
  &\ge \left(q(1-p)-p\right)\rho_{\mathrm{sum}}-(1+q)(1-p)^2\frac{\rho^{2}_{\mathrm{sum}}}{X_{\max}+1}.
  \end{aligned}
\end{equation*}
That is when
\[
\rho_{\mathrm{sum}}< \frac{q(1-p)-p}{(1+q)(1-p)^2}(X_{\max}+1).
\]
We can state that the traffic flow will tend to be free-flow.

The legitimacy of the above boundaries can be easily verified.

\subsubsection{Gaussian distribution hypothesis}

According to empirical data, the speed obeys Gaussian distribution in the equilibrium state after full interaction. We assume that a localized equilibrium state is reached at each $x$. So given $x$, $f(x,v)$ obeys Gaussian distribution i.e.
\begin{equation}\label{11}
f(x,v)\sim N(\mu(x),\sigma^2)\quad \forall x.
\end{equation}

Combining \ref{8} and \ref{11}, we obtain
\begin{equation}\label{12}
\footnotesize
A-D=\left(q(1-p)-p\right)\rho_{\mathrm{sum}}-(1+q)(1-p)^2 \frac{1}{2\pi\sigma^2}\sum_{v}\left(\sum_{x=0}^{X_{\max}}\sum_{\tau=0}^{v}exp\left(-\frac{(\tau-\mu(x))^2+(v-\mu(x))^2}{2\sigma^2}\right)f^{t}(x)f^{t}(x)\rho^{2}_{\mathrm{sum}}\right).
\end{equation}
\begin{equation*}
  \small
  \begin{aligned}
  A-D&\le \left(q(1-p)-p\right)\rho_{\mathrm{sum}}- (1+q)(1-p)^2 \frac{1}{2\pi\sigma^2}exp\left(-\frac{2(V_{\max}+1)^2}{2\sigma^2}\right)\sum_{v}\sum_{\tau=0}^{v}\sum_{x=0}^{X_{\max}}f^{t}(x)f^{t}(x) \rho^{2}_{\mathrm{sum}},\\
  &\le \left(q(1-p)-p\right)\rho_{\mathrm{sum}}-(1+q)(1-p)^2 \frac{1}{2\pi\sigma^2}exp\left(-\frac{2(V_{\max}+1)^2}{2\sigma^2}\right)\frac{V_{\max}^2}{2} \frac{\rho^{2}_{\mathrm{sum}}}{X_{\max}+1}.
  \end{aligned}
\end{equation*}
where the second inequality holds because

\begin{equation*}
  \begin{aligned}
 \sum_{x=0}^{X_{\max}}f^{t}(x)f^{t}(x) & =  \left(\sum_{x=0}^{X_{\max}}f^{t}(x)\right)^2-2\sum_{i=0}^{X_{\max}}\sum_{j=i+1}^{X_{\max}}f^{t}(i)f^{t}(j),\\
 \left(\sum_{x=0}^{X_{\max}}f^{t}(x)\right)^2&\le (X_{\max}+1)  \sum_{x=0}^{X_{\max}}f^{t}(x)f^{t}(x)  \quad (ab\le \frac{a^2+b^2}{2}).\\
 \sum_{x=0}^{X_{\max}}f^{t}(x)f^{t}(x)& \ge \frac{1}{X_{\max}+1}.
  \end{aligned}
\end{equation*}

So when
\[
\rho_{\mathrm{sum}}>2(X_{\max}+1)2\pi\sigma^2\frac{q(1-p)-p}{(1+q)(1-p)^2 exp\left(-\frac{2(V_{\max}+1)^2}{2\sigma^2}
  \right)V_{\max}^2},
\]
The whole system will have more cars accelerating in all $t$ cases. We can say that in this case the traffic flow will tend to be congested.

Similarly we can have
\begin{equation*}
  \begin{aligned}
  A-D&\ge \left(q(1-p)-p\right)\rho_{\mathrm{sum}}- (1+q)(1-p)^2 \frac{1}{2\pi\sigma^2}\sum_{v}\sum_{\tau=0}^{v}\sum_{x=0}^{X_{\max}}f^{t}(x)f^{t}(x) \rho^{2}_{\mathrm{sum}},\\
  &\ge \left(q(1-p)-p\right)\rho_{\mathrm{sum}}-(1+q)(1-p)^2 \frac{1}{2\pi\sigma^2}V_{\max}^2\rho^{2}_{\mathrm{sum}}.
  \end{aligned}
\end{equation*}
That is when
\[
\rho_{\mathrm{sum}}<2\pi\sigma^2\frac{q(1-p)-p}{(1+q)(1-p)^2 V_{\max}^2},
\]
We can state that the traffic flow will tend to be free-flow.

We can see that when the density is large enough it spontaneously will enter the congested flow, and when the density is small enough it will tend to free flow on the contrary. In the next section we will select different parameters for numerical simulations.
\section{Numerical Simulation}

Our numerical simulation is divided into the following parts. First, comparing the effects of different densities and different initial distributions on the final traffic flow steady state under two assumptions; second, comparing the difference and consistency of phase transitions between the two assumptions.

In the numerical experiments, our parameter design follows empirical data and general knowledge. The random diffusion probability usually takes values in the range $0.15-0.2$, and in the following numerical simulation, we set $p=0.15$. And the acceleration probability is set to\cite{7}
\[
q=(1-\rho_{sum})^2.
\]
This could reflect the fact that the higher the density, the lower the acceleration probability, and the lower the density, the higher the acceleration probability.

Roads divided into 2000 cells with periodic boundary conditions. Speed is divided into 100 cells, with the lowest speed being 0. Let the time step be 1000 iterations.

\subsection{Homogenization}

\begin{figure}[!ht]
\centering
\begin{minipage}{0.3\textwidth}
  \includegraphics[width=4.5cm]{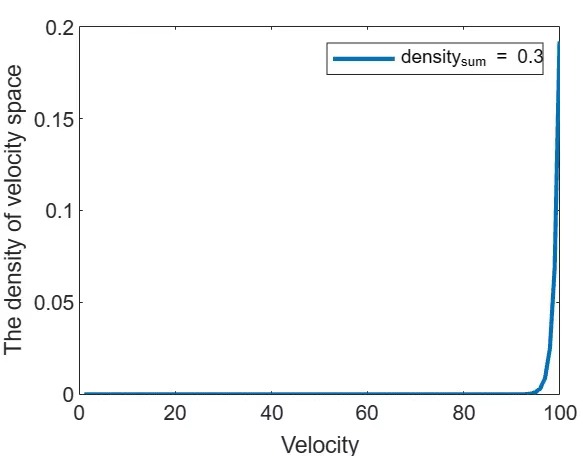}
  \caption*{\small Free-flow state}
\end{minipage}%
\begin{minipage}{0.3\textwidth}
  \includegraphics[width=4.7cm]{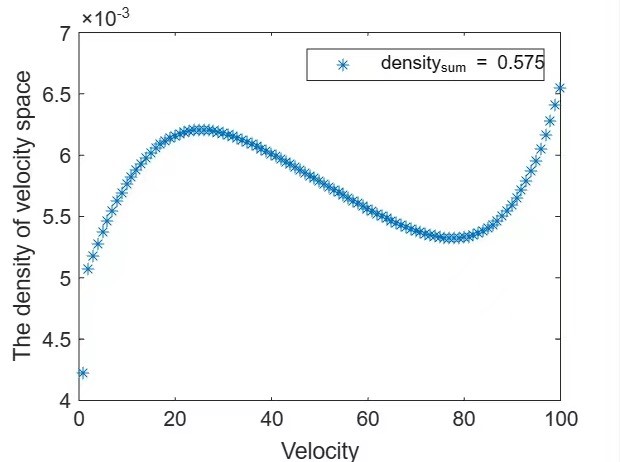}
  \caption*{\small Phase transition point}
\end{minipage}%
\begin{minipage}{0.3\textwidth}
  \includegraphics[width=4.5cm]{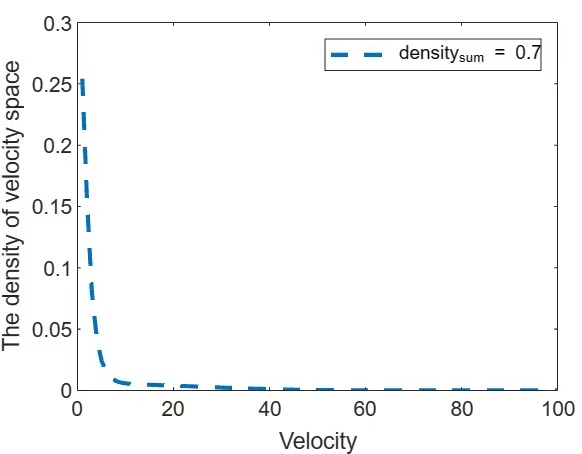}
  \caption*{\small Congested state}
\end{minipage}
\caption{The case of spatial homogenization: the effect of different densities on the stability point of $v$-space.}
\end{figure}

From Figure 2, we can clearly see the phenomenon of phase change. When the density is small, the vast majority of vehicles traveling at nearly the highest speed; when the density is in the vicinity of 0.575, here is the phase transition point, the vehicle speed distribution is more evenly distributed; when the density is larger, the vehicle tends to be basically congested state.

It is important to note:
\begin{itemize}
  \item 1. The acceleration probability and the random diffusion probability affect the location of the phase transition point, but not the existence of the phase transition phenomenon. Our numerical simulation parameters are chosen from empirical data and common parameter settings recognized in popular research.
  \item 2. The assumption of spatial homogenization facilitates congestion evacuation because it forces low-speed vehicles to be evenly distributed throughout the space. So for the real case, the density corresponding to the phase transition point should be less than the density corresponding to the spatially homogenized phase transition point.
\end{itemize}

In the following, we will experiment whether the initial conditions have an impact on the final converged state. For $\rho_\mathrm{sum}=0.3,0.575,0.7$, we experimented with three initial conditions: Initial-Low: the density is initially uniformly distributed throughout the first quintile of the velocity region, i.e., the low-speed region; Initial-Uniform: the density is initially uniformly distributed throughout the whole velocity space; Initial-High: the density is initially uniformly distributed throughout the back quintile of the velocity space, i.e., the high-speed region.

\begin{figure}[!ht]
\centering
\begin{minipage}{0.3\textwidth}
  \includegraphics[width=4.7cm]{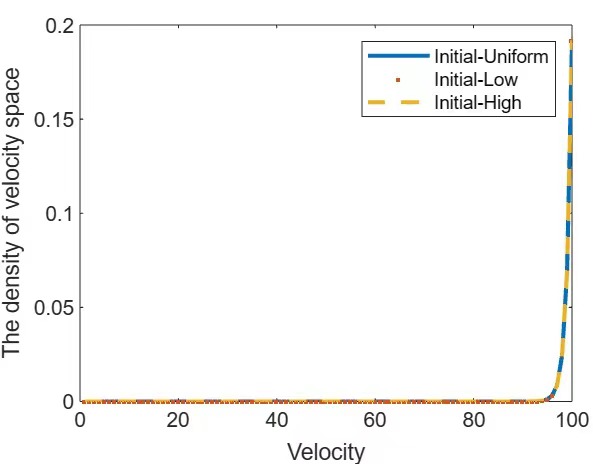}
  \caption*{\small $\rho_\mathrm{sum}=0.3$}
\end{minipage}%
\begin{minipage}{0.3\textwidth}
  \includegraphics[width=4.7cm]{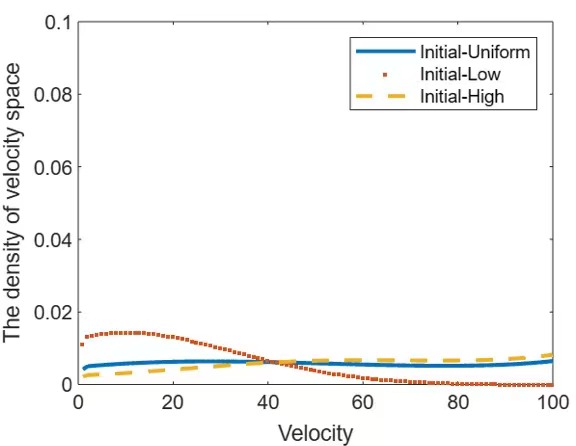}
  \caption*{\small $\rho_\mathrm{sum}=0.575$}
\end{minipage}%
\begin{minipage}{0.3\textwidth}
  \includegraphics[width=4.7cm]{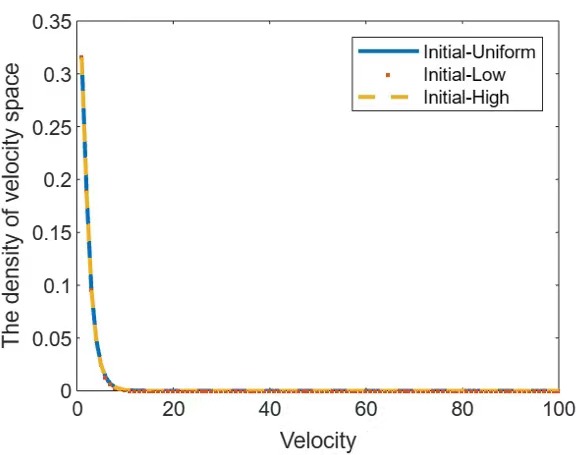}
  \caption*{\small $\rho_\mathrm{sum}=0.7$}
\end{minipage}
\caption{The case of spatial homogenization: the effect of different initial distributions of the same density on the $v$-space stabilization point}
\end{figure}

From the figure 3, it can be seen that the initial conditions have no effect on the final steady state for both high and low density. For densities in the vicinity of the phase transition point, the initial conditions cause a perturbation, but the overall velocity distribution is uniform. 

\subsection{Gaussian assumption}

We mentioned in the previous subsection that spatial homogenization can make traffic flow conducive to evacuation, and in this subsection we circumvent this drawback by assuming that the density of vehicles in the geometric space obeys a single peaked curve. The case of multiple peaks can be decomposed into multiple single-peak cases to be handled. For simplicity, we use a Gaussian bell curve to describe this single-peaked curve.

Empirical data show that the velocity follows a Gaussian distribution\cite{8}, so we assume that for given $x$
\[
f(x,v)\sim N(\mu(x),\sigma_{v}^{2}),
\]
where
\[
\mu(x)=V_{max}(1-\frac{\rho(x)}{\rho_{max}})= V_{max}(1-f(x)\rho_{\mathrm{sum}}),
\]
and assume $\sigma_v^2 = 0.2\times V_{max}$.

Density increases near congestion points, so geometric density exhibits peaks in the roadway. On the one hand, multiple congestion points merge or dissipate over a longer evolutionary process; on the other hand, multiple peaks can be considered as a composite function of a single peak function, so we assume that the geometric density obeys a Gaussian distribution:
\[
f(x)\sim N(\frac{X_{\max+1}}{2},\sigma_x^2),
\]
where the mean value is $\frac{X_{\max+1}}{2} $ because of the periodic boundary condition and the variance can be  estimated using relative entropy.
\begin{equation*}
  D_{KL}(P\mid Q)=\sum P(i)\log(\frac{P(i)}{Q(i)}).
\end{equation*}
where $P$ and $Q$ are two probability distributions, and the relative entropy is equal to 0 only when $P=Q$.

 The use of relative entropy is justified here because we want to make the distribution obtained by the hypothesis as close as possible to the real evolved distribution.

From equation 3, we need to utilize $f^{t}(x,v)$ to get $f^{t+1}(v)$ during the time update process. The spatial homogenization $f^{t}(x,v)=\frac{f^{t}(v)}{X_{\max+1}}$ defines the updating process in $v$-space. We avoid the limitation of spatial homogenization by the following idea. See figure 4.

\begin{figure}[!h]
\centering
\begin{tikzpicture}[node distance=2cm]
  \node (A) {$f^{t}(x)=\frac{1}{\sqrt{2\pi}\sigma} e^{-\frac{(x-(X_{\max}+1)/{2})^2}{2\sigma^2}}$};
  \node (B) [left of=A,node distance=3cm] {$f^{t}(v)$};
  \node (C) [below of=A] {$\hat{f}^{t}(x,v)$};
  \node (D) [below of=C] {$\hat{f}^{t}(v)$};
  \node (E) [below of=D] {$f^{t}(x,v)$};
  \node (F) [right of=E,node distance=5cm] {$f^{t+1}(v)$};

  \draw[->] (A) to node[right] {$\hat{f}^{t}(x,v) = \frac{1}{\sqrt{2\pi}\sigma_v} e^{-\frac{(v-\mu(x))^2}{2\sigma_v^2}}$} (C);
  \draw[->] (C) to node[right] {$\hat{f}^{t}(v) = \sum_{x=0}^{X_{\max}}\hat{f}^{t}(x,v)$} (D);

  \draw[->] (B) to  node[left] {Using relative entropy to find $\sigma$} (D);
  \draw[->] (D) to  node[right] {Utilizing $\sigma $ yields $f^{t}(x,v)$} (E);
  \draw[->] (E) to node[above] {Updating process}(F);
\end{tikzpicture}
\caption{\small $\hat{f}$ denotes the result derived from the hypothesis, and then the relative entropy is used to obtain the value of $\sigma$ using the real evolved result $f^{t}(v)$, which leads to $f^{t}(x,v)$, which is further evolved to $f^{t+1}(v)$}
\end{figure}
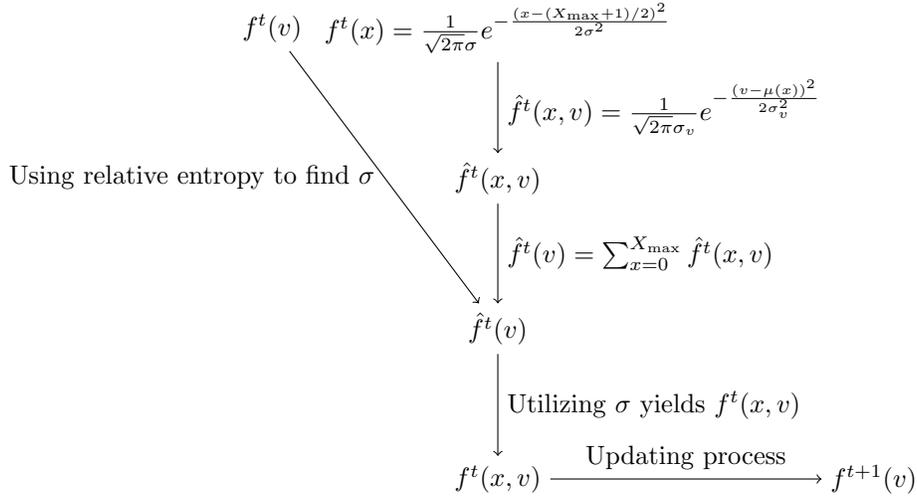

Under the assumption of Gaussian distribution, we test the difference in the final steady state of the traffic flow at different densities.

\begin{figure}[!h]
\centering
\begin{minipage}{0.3\textwidth}
  \includegraphics[width=4.5cm]{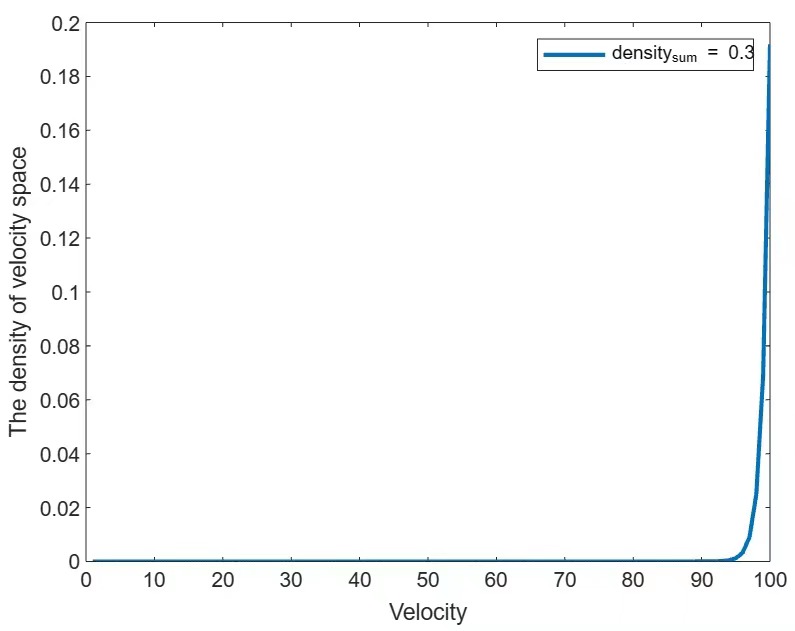}
  \caption*{\small Free-flow state}
\end{minipage}%
\begin{minipage}{0.3\textwidth}
  \includegraphics[width=4.5cm]{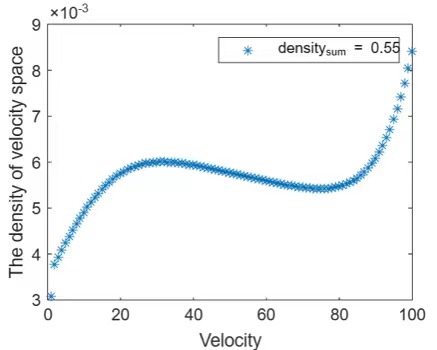}
  \caption*{\small Phase transition point}
\end{minipage}%
\begin{minipage}{0.3\textwidth}
  \includegraphics[width=4.5cm]{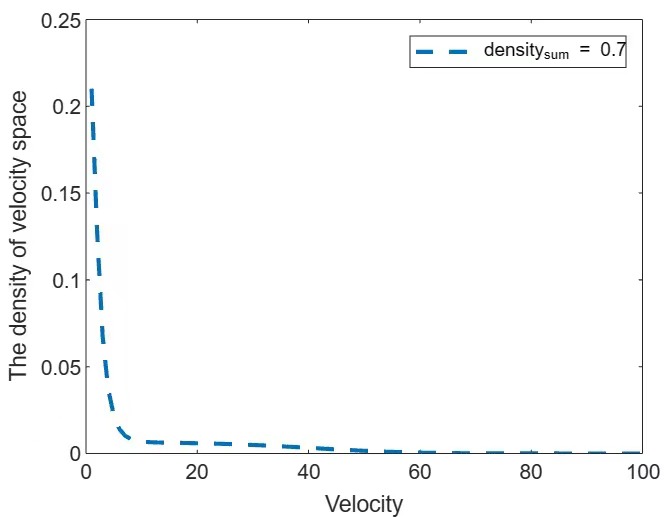}
  \caption*{\small Congested state}
\end{minipage}
\caption{The case of Gaussian assumption: the effect of different densities on the stability point of $v$-space.}
\end{figure}

From Figure 5, we can clearly see the phenomenon of phase change. At low densities, the vast majority of cars travel at high speeds, and at high densities, the vast majority of cars travel at low speeds. We then test the effect of the initial distribution on its final stabilization under the Gaussian assumption. The initial distribution is consistent with the homogenization case, considering three cases where the initial velocity is located in the low-speed region, the full space, and the high-speed region.

\begin{figure}[!ht]
\centering
\begin{minipage}{0.3\textwidth}
  \includegraphics[width=4.8cm]{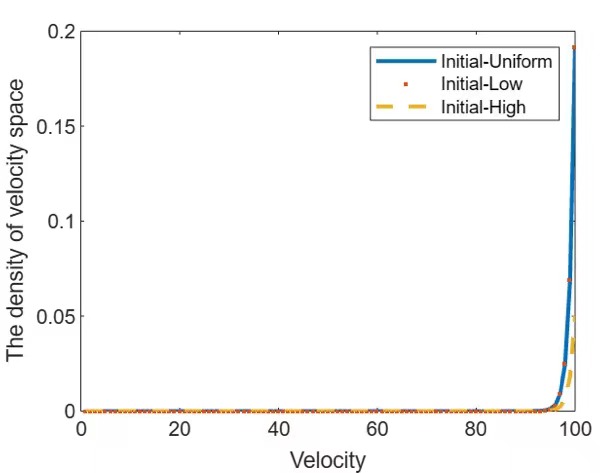}
  \caption*{\small $\rho_\mathrm{sum}=0.3$}
\end{minipage}%
\begin{minipage}{0.3\textwidth}
  \includegraphics[width=4.8cm]{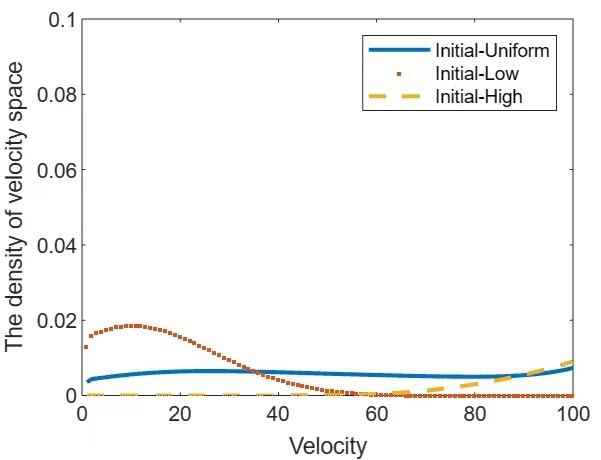}
  \caption*{\small $\rho_\mathrm{sum}=0.525$}
\end{minipage}%
\begin{minipage}{0.3\textwidth}
  \includegraphics[width=4.8cm]{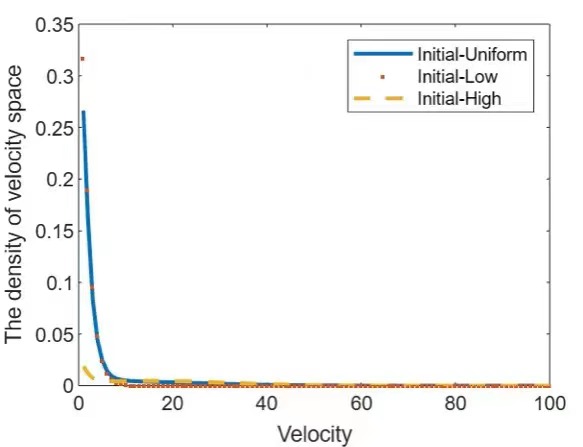}
  \caption*{\small $\rho_\mathrm{sum}=0.7$}
\end{minipage}
\caption{The case of Gaussian assumption: the effect of different initial distributions of the same density on the $v$-space stabilization point}
\end{figure}

Similar to the case of spatial homogenization, from the figure 6, we can see that the initial distribution has essentially no effect on the overall trend. The initial distribution has a slight effect on the final steady state. Compared to an initial distribution with a uniform distribution over the whole space, an initial speed in the low-speed region results in a slightly "congested" final state, and an initial speed in the high-speed region results in a slightly "free" final result.

We compare the stability of the two models at the same density.

\begin{figure}[!h]
\centering
\begin{minipage}{0.5\textwidth}
  \includegraphics[width=6.7cm]{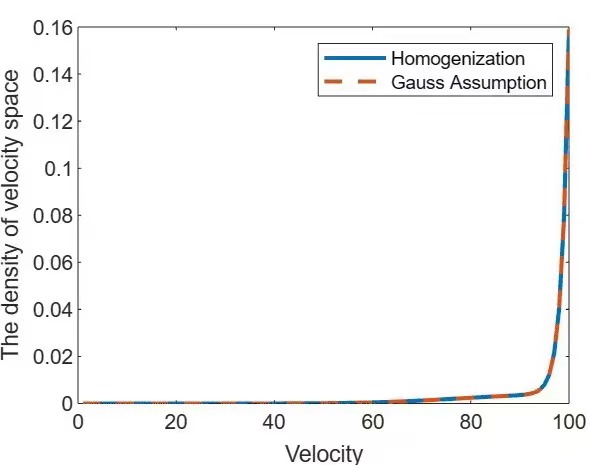}
  \caption*{\small $\rho_\mathrm{sum}=0.3$}
\end{minipage}%
\begin{minipage}{0.5\textwidth}
  \includegraphics[width=7cm]{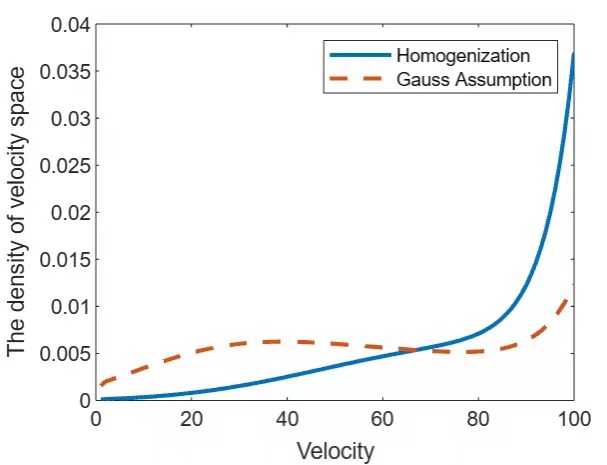}
  \caption*{\small $\rho_\mathrm{sum}=0.55$}
\end{minipage}
\caption{Comparison of results for Gaussian assumptions and spatial homogenization.}
\end{figure}

Simulations have done under the same density and initial distribution. From the figure 7, we can see in the state of extreme freedom, the curves of the two models are essentially overlapping. But with $\rho_\mathrm{sum} = 0.55$, the spatial homogenization will lead to slightly more "free" states in the final stable state than the Gaussian assumption. This fits with our theoretical analysis -- spatial homogenization accelerates the evaporation of congestion.
\section{Conclusion}
The Boltzmann-type equation describes the overall behavior of a system when there are multiple agents interacting. However, when applying it to traffic flow, the continuous equation in the thermodynamic limit produce a non-negligible difference from the real situation, since the size of a vehicle is sizeable compared to a section of road. To address this issue, we model it in the context of discrete space, discrete time and discrete speed. In contrast to the common cellular automata, it still discusses the way the density in phase space changes with time, rather than modeling the dynamics of each agent. So the problem of noise from a single simulation is avoided. However, we can still maintain the flexibility of the cellular automata, for example, effects such as stochastic slowing can be easily introduced into our model. We assume that the density of vehicles in space is uniform or single-peaked, and solve the evolutionary trend of the speed distribution, and explaining the phase transition in speed as the density of vehicles changes. This model can be relied upon to research how the multiple parameters will affect the dynamic outcome of the traffic flow. Afterwards, new treatments may also be developed to directly solve the complex dynamics in phase space.

\newpage
\bibliographystyle{plain}

\end{document}